\newcommand{\Ir}{Z\!\!\!Z}
\newcommand{\Ibb}[1]{ {\rm I\ifmmode\mkern
            -3.6mu\else\kern -.2em\fi#1}}
\newcommand{\ibb}[1]{\leavevmode\hbox{\kern.3em\vrule
     height 1.2ex depth -.3ex width .2pt\kern-.3em\rm#1}}
\newcommand{\Cx}{{\ibb C}}
\newcommand{\Rl}{{\Ibb R}}
\documentstyle[12pt]{article}
\begin{document}
\renewcommand{\theequation} {\arabic{section}.\arabic{equation}}
\begin{tabbing}
\hspace*{12cm}\= GOET-TP 55/93 \\
              \> October 1993
\end{tabbing}
\vspace*{2cm}
\centerline{\huge \bf Noncommutative Differential Calculus:}
\vskip.5cm
\centerline{\Large \bf Quantum Groups, Stochastic Processes, and the
                      Antibracket}
\vskip1.5cm

\begin{center}
      {\bf Aristophanes Dimakis}
      \vskip.1cm
      Department of Mathematics, University of Crete \\
      GR-71409 Iraklion, Greece
      \vskip.2cm
       and
      \vskip.2cm
      {\bf Folkert M\"uller-Hoissen}
      \vskip.1cm
      Institut f\"ur Theoretische Physik  \\
      Bunsenstr. 9, D-37073 G\"ottingen, Germany
\end{center}

\begin{abstract}
\noindent
We explore a differential calculus on the algebra of
$C^\infty$-functions on a manifold. The former is `noncommutative' in
the sense that functions and differentials do not commute, in general.
Relations with bicovariant differential calculus on certain quantum
groups and stochastic calculus are discussed. A similar differential
calculus on a superspace is shown to be related to the
Batalin-Vilkovisky antifield formalism.
\end{abstract}

\renewcommand{\theequation} {\arabic{section}.\arabic{equation}}

\section{Introduction}
\setcounter{equation}{0}
Since Connes' work on noncommutative geometry, the notion of
differential calculus on algebras has entered the realm of physics
through numerous publications. As the commutative algebra of
($\Cx$-valued) functions on a topological space carries
all the information about the space in its algebraic structure,
certain noncommutative algebras may be regarded as a generalization
of the notion of a `space'. If the algebra $\cal A$ is associative,
one can enlarge it to a differential algebra,
a kind of analogue of the algebra of differential forms on a
differentiable manifold.

More precisely, this is a $\Ir$-graded associative algebra
$\bigwedge({\cal A}) = \bigoplus_{r \geq 0} \bigwedge^r({\cal A})$
where $\bigwedge^0 = \cal A$. The spaces $\bigwedge^r({\cal A})$
of {\em $r$-forms} are generated as $\cal A$-bimodules via the action of
an {\em exterior derivative} $d \, : \, \bigwedge^r({\cal A})
\rightarrow \bigwedge^{r+1}({\cal A})$ which is a linear operator
acting in such a way that $d^2 =0$ and $d(\omega \omega') = (d\omega)
\, \omega' + (-1)^r \omega \, d\omega'$ (where $\omega$ and $\omega'$
are $r$- and $r'$-forms, respectively). Without further restrictions,
$\bigwedge({\cal A})$ is the so-called {\em universal differential
envelope} of $\cal A$. It associates, for example, independent
differentials with $f \in \cal A$ and $f^2$.

What we would rather
like to have is a closer analogue of the algebra of differential
forms on a manifold. In particular, if $\cal A$ is generated by
a set of $n$ elements (e.g., coordinate functions $x^i$ on a
manifold), we might want the space of 1-forms to be generated as
a left- (or right-) $\cal A$-module by the differentials $dx^i$.
In order to achieve this, one
has to add commutation rules for functions and differentials to
the differential algebra structure defined above. In case of the
commutative algebra of $C^\infty$-functions on a manifold,
the ordinary calculus of differential forms simply assumes
that 1-forms and functions commute. If, however, $\cal A$ is the
algebra of functions on a discrete set, this assumption cannot
be kept. The algebra of functions on a two-point set, for example,
is generated by a function $y$ such that $y^2=1$. Acting with $d$
on this relation yields $y \, dy = - dy \, y$ and thus
{\em anti}-commutativity.\footnote{In this example the commutation
relation is {\em not} an additional assumption, but follows from the
general rules of differential calculus. This is a special feature of
the two-point space.} This example plays a crucial role in
models of elementary particle physics \cite{Conn+Lott}. Here we
just take it to illustrate what we mean by `noncommutative
differential calculus', namely noncommutativity between functions
and differentials.

Let $\cal A$ be the set of functions on $\Rl$ generated by a
coordinate function $x$ (and a unit element which we identify
with $1 \in \Cx$). The simplest consistent deformation of the ordinary
differential calculus is then determined by $\lbrack x , dx \rbrack =
a \, dx$ where $a$ is a positive real constant. If we define
partial derivatives by $df = \stackrel{\leftarrow}{\partial}\!\!\! f
\, dx = dx \! \stackrel{\rightarrow}{\partial} \!\!\! f$, they turn out
to be (left- and right-) {\em discrete} derivatives. An integral is
naturally associated with $d$ and (for the higher-dimensional
generalization of the calculus) it turns out that the deformation
from $a=0$ to $a>0$ transforms continuum theories (like a gauge
theory) to the corresponding lattice theory (where $a$ plays the
role of the lattice spacing) \cite{DMHS93-latt}. A simple coordinate
transformation brings the above commutation relation into the form
$y \, dy = q \, dy \, y$ with $q \in \Cx$, the differential calculus
underlying $q$-calculus \cite{DMH92-q}.
This noncommutative differential calculus is the best understood
and most complete example so far. We can also introduce it
on the space of functions on a lattice with
spacings $a$ instead of $\cal A$. More generally, differential
calculus on discrete sets is supposed to be of relevance for approaches
towards discrete field theory and geometry (see \cite{DMH93-finset} and
the references given there).

Another interesting example of a noncommutative differential
calculus on a commutative algebra is the following
\cite{DMH92-grav,DMH93-stoch}.
Let $\cal A$ be the algebra of $C^\infty$-functions on a manifold
$\cal M$ and let us assume the following commutation relations expressed
in terms of local coordinates $x^i$:
\begin{eqnarray}        \label{x-dx}
   \lbrack x^i , dx^j \rbrack = \gamma \, g^{ij} \, dt
\end{eqnarray}
where $\gamma$ is a constant, $g$ a real symmetric tensor (e.g., a
metric) on $\cal M$, and $t$ an `external' (time) parameter.
The above commutation relation is actually coordinate independent.
The differential calculus based on it is related to quantum mechanics
\cite{DMH92-grav} and stochastics \cite{DMH93-stoch}
(depending on whether $\gamma$ is imaginary or real), and to `proper
time' (quantum) theories \cite{DMH92-grav}.
A generalization of (\ref{x-dx}) is obtained by replacing
$\gamma \, dt$ by a 1-form $\tau$, i.e.
\begin{eqnarray}       \label{x-dx-tau}
           [x^i , dx^j] = \tau \, g^{ij}
\end{eqnarray}
where $\tau$ should have the following properties,
\begin{eqnarray}       \label{tau-rels}
   [x^i , \tau] = 0  \quad , \quad
   \tau \, \tau = 0  \quad , \quad d\tau = 0    \; .
\end{eqnarray}
This structure in fact shows up in the classical limit ($q \to 1$)
of (bicovariant \cite{Woro89}) differential calculus on certain quantum
groups \cite{MH+Reut93}. For functions $f,h \in \cal A$, we have
\begin{eqnarray}             \label{(f,h)_g}
           [f , dh] = \tau \, (f,h)_g  \quad , \quad
           (f,h)_g := g^{ij} \, \partial_i f \, \partial_j h
\end{eqnarray}
(where $\partial_i := \partial/\partial x^i$). In sections 2-5, a brief
introduction to various aspects of this differential calculus is given.
Some of the results, in particular in sections 3 and 5, have not been
published before.

Sections 6 and 7 present basically new results. We
introduce a differential calculus on a superspace and show that the
antibracket and the $\Delta$-operator of the Batalin-Vilkovisky
formalism \cite{BV} (developed for quantization of gauge theories)
appear naturally in this framework. A corresponding generalization of
gauge theory is also formulated. The differential calculus is a kind of
superspace counterpart of the abovementioned differential calculus on
manifolds.

Our work establishes relations between noncommutative differential
calculus and various mathematical structures which play a role in
physics. The latter are thus put into a new perspective
which will hopefully contribute to an improved understanding and
handling of these structures.

\section{The classical limit of bicovariant differential calculi on
the quantum groups $GL_q(2)$ and $SL_q(2)$}
\setcounter{equation}{0}
Let us denote the entries of a $GL(2)$-matrix as follows,
\begin{eqnarray}
   M =  \left(\begin{array}{cc} x^1 & x^2    \\
                                x^3 & x^4
              \end{array}\right)             \; .
\end{eqnarray}
Let $\cal A$ be the algebra of polynomials in $x^i$. The quantum group
$GL_q(2)$ is a noncommutative deformation of $\cal A$ as a Hopf
algebra. The structure of a quantum group allows to narrow down the
many possible differential calculi on it. This results in the notion
of {\em bicovariant} differential calculus \cite{Woro89}. For
$GL_q(2)$ there is a 1-parameter set of bicovariant differential
calculi. In the classical limit $q \to 1$ they lead
\cite{MH+Reut93,DMH93-stoch} to the commutation relations
(\ref{x-dx-tau}) with
\begin{eqnarray}
 g^{ij} &=& (\det M)^{-1} \, x^i \, x^j
            + 4 \, (\delta^{(i}_2 \, \delta^{j)}_3 -
            \delta^{(i}_1 \, \delta^{j)}_4)   \label{GL-g} \\
 \tau &=& s \, (dx^1 \, x^4 - dx^2 \, x^3 - dx^3 \, x^2 + dx^4 \, x^1)
                                              \label{GL-tau}
\end{eqnarray}
where $s$ is a free parameter. The ordinary differential calculus on
$GL(2)$ is only obtained when $s=0$.

The condition for the matrix $M$ to be in $SL(2)$ is the quadratic
equation
\begin{eqnarray}               \label{det=1}
      \det M = x^1 \, x^4 - x^2 \, x^3 = 1  \; .
\end{eqnarray}
Compatibility of the analogous condition for the quantum group $SL_q(2)$
with bicovariant differential calculus restricts the parameter $s$ to
only two values (both different from zero) \cite{MH+Reut93}.
There are thus only two
bicovariant differential calculi on $SL_q(2)$ and for both the
classical limit is not the ordinary differential calculus. We will
only consider one of them here.
In a cordinate patch where $x^1 \neq 0$ we can use $x^a, \, a=1,2,3$,
as coordinates. The differential calculus is then determined by
(\ref{x-dx-tau}) with
\begin{eqnarray}
  g^{ab} &=& x^a \, x^b + 4 \, \delta^{(a}_2 \, \delta^{b)}_3  \\
    \tau &=& {1 \over 3} \, (dx^1 \, x^4 - dx^2 \, x^3
             - dx^3 \, x^2 + dx^4 \, x^1)
\end{eqnarray}
where $x^4 = (1+ x^2 x^3)/x^1$. Although we only have {\em three}
independent coordinates in this case, the space of 1-forms (as a left
or right $\cal A$-module) is {\em four}-dimensional since $\tau$ cannot
be expressed as $\tau = \sum_{a=1}^3 dx^a \, f_a$ with $f_a \in \cal
A$. What's going on here is explained in more detail in the following
section, using a simple example.

\section{Differential calculi on quadratic varieties}
\setcounter{equation}{0}
Let $x^i,\;i=1,\ldots,n$, be real variables, $\alpha_{ij}$ a
nondegenerate symmetric constant form with inverse $\alpha^{ij}$.
We want to construct a noncommutative differential calculus with
(\ref{x-dx-tau}) and (\ref{tau-rels}), compatible
with the quadratic relation
\begin{eqnarray}
      \alpha_{ij} \, x^i \, x^j = 1   \; .       \label{qua}
\end{eqnarray}
The $SL(2)$-condition (\ref{det=1}) provides us with a particular
example. Acting with $d$ on (\ref{qua}) and using (\ref{x-dx-tau}),
we obtain
\begin{eqnarray}
  \tau = dx^i \, (- 2 \, \alpha_{ij} \, x^j/a) =: dx^i \, \tau_i
\end{eqnarray}
where we have assumed that $a:=\alpha_{ij} \, g^{ij} \neq 0$.
The condition $[x^i , \tau] = 0$ implies
\begin{eqnarray}
           g^{ij} \, \tau_j = 0  \; .
\end{eqnarray}
It is natural to look for an expression for $g^{ij}$ in terms of
$\alpha^{ij}$ and the coordinates $x^i$. We are then led to the
following solution of the last equation:
\begin{eqnarray}
     g^{ij} = x^i \, x^j - \alpha^{ij}  \; .
\end{eqnarray}
{}From this we find $a= 1-n$. In the $SL(2)$ case, we
recover (\ref{GL-g}) and (\ref{GL-tau}) with the correct restriction
on the parameter, i.e. $s=1/3$.
\vskip.2cm
\noindent
{\em Example:}
Consider two variables $x,y$ subject to the quadratic relation
\begin{eqnarray}
                x \, y = 1   \; .
\end{eqnarray}
We thus have $n=2$, $\alpha_{ij}=(1/2)(\delta_{i1} \delta_{j2}
 + \delta_{i2}\delta_{j1})$ and
\begin{eqnarray}
    (g^{ij})= \left(\begin{array}{cc}  x^2 & -1    \\
                                       -1  & y^2
                    \end{array}\right)  \; .
\end{eqnarray}
Furthermore, $\tau = dx \, y + dy \, x$. In the case under
consideration, (\ref{x-dx-tau}) is a system of four equations.
Three of them are redundant, however, since they are consequences
of
\begin{eqnarray}                \label{x-dx-ex}
        \lbrack x , dx \rbrack = \tau \, x^2  \; .
\end{eqnarray}
Although we have
only one free coordinate ($x$), the 1-forms $dx$ and $\tau$ are
independent in the sense that $\tau = dx \, (1/x) - (1/x) \, dx$
cannot be expressed as $f(x) \, dx$ or $dx \, f(x)$. The space of
1-forms is therefore two-dimensional (as a left or right
$\cal A$-module, where $\cal A$ is now the algebra of functions of $x$).
We can use the expression for $\tau$ to eliminate $\tau$ from
(\ref{x-dx-ex}). This results in the equation
$ x \, dx - 2 \, dx \, x + (1/x) \, dx \, x^2 = 0 $ which is
insufficient to transform the $\cal A$-bimodule of 1-forms into a left
(or right) $\cal A$-module.

\section{A generalized gauge theory and `second order differential
geometry'}
\setcounter{equation}{0}
It is rather straightforward to formulate a generalization of gauge
theory and differential geometry using the `deformed' differential
calculus on ${\cal A} = C^\infty({\cal M})$ with (\ref{x-dx-tau})
and (\ref{tau-rels}) (see also \cite{DMH92-grav}). It should be
noticed, however, that -- as a consequence of the deformation --
the differential of a function $f$ is now given by
\begin{eqnarray}    \label{df-tau}
    df = \tau \, {1 \over 2} \, g^{ij} \, \partial_i \partial_j f
         + dx^i \, \partial_i f
\end{eqnarray}
and involves a {\em second} order differential operator. If a
(space-time) metric is given, it is natural to identify it with
$g^{ij}$.

Let $\psi$ be an element of ${\cal A}^n$ which transforms as
$\psi \mapsto \psi' = U \, \psi$ under a representation of a Lie
group $G$. For local transformations we can construct a covariant
derivative in the usual way,
\begin{eqnarray}
     D \psi = d \psi + A \, \psi     \; .
\end{eqnarray}
This is indeed covariant if the 1-form $A$ transforms according
to the familiar rule
\begin{eqnarray}             \label{A-trans}
     A' = U \, A \, U^{-1} - dU \, U^{-1}  \; .
\end{eqnarray}
In the following we will only consider the case where the coordinate
differentials $dx^i$ and the 1-form $\tau$ are linearly independent
and form a basis of the space of 1-forms (as a left or right $\cal
A$-module). $A$ can then be written in a unique way as
\begin{eqnarray}
     A = \tau \, {1 \over 2} \, A_\tau + dx^i \, A_i  \; .
\end{eqnarray}
Inserting this expression in (\ref{A-trans}), we find that $A_i$
behaves as an ordinary gauge potential and
\begin{eqnarray}
     A_\tau = g^{ij} \, (\partial_i A_j - A_i A_j) + M
\end{eqnarray}
where $M$ is an arbitrary tensorial part ($M' = U M U^{-1}$).
Since $U$ depends on $x^i$, in general, it does not commute with
$dx^j$.
It is convenient to introduce the gauge-covariant differential
$Dx^i := dx^i - \tau \, A^i$.
The covariant derivative of $\psi$ can now be written as
\begin{eqnarray}
     D \psi = \tau \, {1 \over 2} \, (g^{ij} D_i D_j + M) \, \psi
              + Dx^i \, D_i \psi
\end{eqnarray}
where $D_i$ denotes the ordinary covariant derivative (with $A_i$).
The field strength of $A$ is
\begin{eqnarray}
     F = dA + A^2 = \tau \, {1 \over 2} \, (D^\ast F - DM)
         + {1 \over 2} \, Dx^i \, Dx^j \, F_{ij}
\end{eqnarray}
where $D^\ast F = dx^i \, D^j F_{ji}$ involves the Yang-Mills
operator (when $g^{ij}$ is identified with the space-time metric).
$F_{ij}$ is the (ordinary) field strength of $A_i$.

If $\tau$ behaves as a scalar and $g^{ij}$ as a contravariant
tensor under coordinate transformations, the defining relations
of our differential calculus -- and in particular (\ref{x-dx-tau}) --
are coordinate independent \cite{DMH92-grav,DMH93-stoch}.
The coordinate differentials $dx^i$ do not transform covariantly,
however, since
\begin{eqnarray}
   d{x'}^k = \tau \, {1 \over 2} \, g^{ij} \partial_i \partial_j
             {x'}^k + d x^\ell \, \partial_\ell {x'}^k
\end{eqnarray}
as a consequence of (\ref{df-tau}).
For a vector field $Y^i$ we introduce a (right-) covariant derivative
\begin{eqnarray}
           D Y^i := d Y^i + Y^j \; {}_j \Gamma^i   \; .
\end{eqnarray}
This is indeed {\em right}-covariant iff the generalized connection
${}_j \Gamma^i$ is given by
\begin{eqnarray}
   {}_j \Gamma^i = \tau \, {1 \over 2} \, \left \lbrack
                   g^{k \ell} ( \partial_k \Gamma^i{}_{j \ell}
                   + \Gamma^i{}_{m k} \Gamma^m{}_{j \ell} ) + M^i{}_j
                                          \right \rbrack
                   + dx^k \, \Gamma^i{}_{jk}
\end{eqnarray}
where $\Gamma^i{}_{jk}$ are the components of an ordinary linear
connection on $\cal M$ and $M^i{}_j$ is a tensor.
Let us introduce the {\em right}-covariant 1-forms
\begin{eqnarray}
 D x^k := d x^k + \tau \, {1 \over 2} \, \Gamma^k{}_{ij} g^{ij}
               \; .
\end{eqnarray}
(\ref{df-tau}) can now be rewritten as
\begin{eqnarray}
    df = \tau \, {1 \over 2} \, g^{ij} \nabla_i \nabla_j f
         + D x^i \; \partial_i f
\end{eqnarray}
where $\nabla_i$ denotes the ordinary covariant derivative.
Also the covariant exterior derivative of $Y^i$ can now be written in
an explicitly right-covariant form,
\begin{eqnarray}
  D Y^i = \tau \, {1 \over 2} \, (g^{k \ell} \nabla_k
               \nabla_\ell Y^i + M^i{}_j \, Y^j)
               + D x^j \; \nabla_j Y^i   \; .
\end{eqnarray}
It is interesting that the (covariant) exterior derivative
of a field contains in its $\tau$-part the corresponding part of the
field equation to which it is usually subjected in physical models.
We refer to \cite{DMH92-grav} for further results.

\section{Stochastic differential calculus}
\setcounter{equation}{0}
When $\tau = \gamma \, dt$ as in (\ref{x-dx}), we may consider
(smooth) functions $f(x^i,t)$ depending also on the parameter $t$.
(\ref{df-tau}) then has to be replaced by
\begin{eqnarray}
  df = dt \, (\partial_t + {\gamma \over 2} \, g^{ij} \, \partial_i
       \partial_j ) \, f + dx^i \, \partial_i f   \; .
\end{eqnarray}
Such a formula is wellknown in the theory of stochastic processes
(It\^o calculus) \cite{Arno74} and suggests that our noncommutative
differential calculus provides us with a convenient framework to
deal with stochastic processes on manifolds. There is indeed a kind
of translation \cite{DMH93-stoch} to the (It\^o) calculus of stochastic
differentials. This can be used to carry the {\em expectation} map
from the latter over to our calculus.
In this section, we introduce an expectation $\bf E$ on the (first
order) differential calculus in a more formal way.
It is then shown for a specific example, that our rules
reproduce familiar results.

Let us consider the equation (\ref{x-dx}) in one dimension (for
simplicity). We write it in the form
\begin{eqnarray}               \label{X_t-comm}
            \lbrack X_t , d X_t \rbrack = dt
\end{eqnarray}
viewing $X_t$ as a process on $\Rl$, a map $\Rl \times \lbrack 0,
\infty ) \rightarrow \Rl$.
$\cal A$ denotes the algebra of smooth functions of $X_t$ and
$t$, and ${\cal F}$ the subalgebra of functions of $t$ only. Let
$\bf E$ be an $\cal F$-linear map ${\cal A} \rightarrow {\cal F}$
which is the identity on $\cal F$. We extend it to 1-forms as
an $\cal F$-linear map via\footnote{On the
rhs of the first equation in (\ref{E-rules}), $d$ is the ordinary
exterior derivative. The second equation can be interpreted by
saying that, given $f_t$, a further increment $dX_t$ is statistically
independent (i.e., $f_t$ is `nonanticipating').
Then, as a consequence of (\ref{X_t-comm}), ${\bf E} (f_t \, dX_t)$
does {\em not} vanish, in general. Here we should view $f_t$ as
evaluated {\em after} a time step $dt$ with increment $dX_t$ in $X_t$.}
\begin{eqnarray}                          \label{E-rules}
 {\bf E} \, df_t = d ({\bf E} f_t)        \quad , \quad
 {\bf E} (dX_t \, f_t) = 0                \qquad
 ( \forall f_t \in {\cal A} ) \; .
\end{eqnarray}
\vskip.2cm \noindent
{\em Example:} (Ornstein-Uhlenbeck process)   \\
Let us consider the differential equation
\begin{eqnarray}          \label{OU}
       dY_t = - k \, dt \, Y_t + \sigma \, dX_t
\end{eqnarray}
with constants $k,\sigma$. For ${\bf E} Y_t$ we obtain from
(\ref{OU}) the ordinary differential equation
\begin{equation}
       d {\bf E} Y_t = - k \, {\bf E} Y_t \, dt
\end{equation}
with the solution ${\bf E} Y_t = {\bf E} Y_0 \, e^{-kt}$. Let us now
show how to calculate higher moments. With
\begin{equation}
   [Y_t , dY_t] = \sigma \, [Y_t , dX_t] = \sigma \,
   [X_t , dY_t] = \sigma^2 \, dt   \; .
\end{equation}
we find
\begin{equation}
 d(Y_t^2) = dY_t \, Y_t + Y_t \, dY_t = 2 \, dY_t \, Y_t
   + \sigma^2 \, dt = 2 \, \sigma \, dX_t \, Y_t
   + dt \, (\sigma^2 - 2 k \, Y_t^2)
\end{equation}
and, using ${\bf E} (dX_t Y_t)=0$, the ordinary differential
equation
\begin{equation}
  d({\bf E} Y_t^2) = dt \, (\sigma^2 - 2 \, k \, {\bf E} Y_t^2)
\end{equation}
for the second moment. The solution is
\begin{equation}
  {\bf E} Y_t^2 = e^{-2 kt} \, {\bf E} Y_0^2 + {\sigma^2 \over 2 k}
  \, (1 - e^{-2 kt})   \; .
\end{equation}
If the moments ${\bf E} Y_0^n$ are given, we obtain in this way the
moments ${\bf E} Y_t^n, \, t > 0$.
The results are the same as if we treat (\ref{OU})
as an (It\^o) stochastic differential equation, which is the
Ornstein-Uhlenbeck equation (see \cite{Arno74}, for example).
We have used rather unusual techniques, however, namely a
noncommutative differential calculus.

\section{A differential calculus on superspace}
\setcounter{equation}{0}
So far we dealt with a commutative algebra generated by coordinate
functions $x^i, \, i=1, \ldots, n$. In this section we enlarge it
to an algebra $\cal A$ of functions on a superspace by adding odd
variables $\xi_i$ and $\eta$. Again, we associate with $\cal A$ a
differential algebra $\bigwedge ({\cal A})$ via the action of an
exterior derivative $d$. In the case of superalgebras a different
version of the Leibniz rule is usually adopted \cite{Kast88},
\begin{eqnarray}           \label{grad-Leibniz}
   d (\omega \, \omega') = d \omega \, \omega' + \hat{\omega} \,
                           d \omega'
\end{eqnarray}
where the hat denotes the grading involution.\footnote{This is defined
on $\bigwedge({\cal A})$ by $\hat{x}^i = x^i, \, \hat{\xi}_i = - \xi_i,
\, \hat{\eta} = - \eta, \, \widehat{d \omega} = -d \hat{\omega}, \,
\widehat{\omega \omega'} = \hat{\omega} \hat{\omega}'$ and linearity.
In particular, the $dx^i$ are odd and $d\eta , \; d\xi_i$ are even.}
In the even sector of $\cal A$, (\ref{grad-Leibniz}) coincides with
our previous rule, however.
We write $\lbrack \, , \, \rbrack$ for the graded commutator (i.e.,
$\lbrack \omega , \omega' \rbrack = \omega \omega' - \omega' \omega$
for $\omega$ even and $\lbrack \omega , \omega' \rbrack =
\omega \omega' - \hat{\omega}' \omega$ for $\omega$ odd).
The universal differential calculus is now restricted by the
following relations,
\begin{eqnarray}
  \lbrack x^i, d\xi_j \rbrack = - \lbrack \xi_j, dx^i \rbrack
                              =  d\eta \, \delta^i_j    \; .
\end{eqnarray}
The remaining graded commutators between superspace coordinates
and their differentials are taken to be zero (so that we have the
standard rules in the pure even and odd sectors).
This defines a consistent differential calculus where the space of
1-forms is generated as a right (or left) $\cal A$-module by
$dx^i, d\xi_j, d\eta$. The differential of a function $f$ on the
superspace can then be expressed as
\begin{eqnarray}                 \label{df}
 df = d\eta \, \tilde{\partial}_\eta f + dx^i \, \tilde{\partial}_i f
      + d\xi_i \, \tilde{\zeta}^i \! f
\end{eqnarray}
where $\tilde{\partial}_\eta, \, \tilde{\partial}_i, \,
\tilde{\zeta}^i$ are operators on $\cal A$.
Using (\ref{grad-Leibniz}) and the basic commutation relations, we find
\begin{eqnarray}      \label{dx-f}
  \lbrack dx^i, f \rbrack = - d\eta \, \tilde{\zeta}^i \! f
  \quad , \quad
  \lbrack d\xi_i, f \rbrack = - d\eta \, \tilde{\partial}_i f  \; .
\end{eqnarray}
With the help of these relations, the Leibniz rule (\ref{grad-Leibniz})
for $d$ now implies
\begin{eqnarray}
 \tilde{\partial}_i (fh) & = & (\tilde{\partial}_i f) \, h +
                               f \, (\tilde{\partial}_i h)
                               \quad , \quad
 \tilde{\zeta}^i (fh) \, = \, (\tilde{\zeta}^i f) \, h +
                            \hat{f}(\tilde{\zeta}^i h)    \\
 \tilde{\partial}_\eta (fh) & = & (\tilde{\partial}_\eta f) \, h
          + \hat{f} \, (\tilde{\partial}_\eta h)
          + (\tilde{\zeta}^i f) \, \tilde{\partial}_i h
          + (\tilde{\partial}^i \hat{f}) \, \tilde{\zeta}_i h  \; .
\end{eqnarray}
Together with $\tilde{\partial}_i x^j = \delta^j_i =
\tilde{\zeta}^j \xi_i, \, \tilde{\partial}_\eta \eta =1$ (a
consequence of (\ref{df})), this leads to
\begin{eqnarray}
 \tilde{\partial}_i = \partial_i := {\partial \over \partial x^i}
     \quad , \quad
 \tilde{\zeta}^i = \zeta^i := {\partial_{(\ell)} \over \partial \xi_i}
     \quad , \quad
 \tilde{\partial}_\eta = \partial_\eta + \Delta := {\partial_{(\ell)}
     \over \partial \eta} + \zeta^i \partial_i
\end{eqnarray}
(where a subscript $(\ell)$ indicates that the derivative is taken
from the left). Hence
\begin{eqnarray}
  df = d\eta \, (\partial_\eta f + \Delta f) + dx^i \, \partial_i f
       + d\xi_i \, \zeta^i \! f  \; .
\end{eqnarray}
Using (\ref{dx-f}), we obtain
\begin{eqnarray}        \label{comm-antibr}
                [f , dh] = d\eta \; (f,h)
\end{eqnarray}
where on the rhs appears the {\em antibracket} \cite{BV}
\begin{eqnarray}
 (f,h) := (\partial_i f) \, \zeta^i h + (\zeta^i \! \hat{f})
            \, \partial_i h
        = \Delta (\hat{f} h) - (\Delta \hat{f}) \, h
          - f \, \Delta h    \; .
\end{eqnarray}
The operator $\Delta$ satisfies $\Delta^2=0$.

The relation (\ref{comm-antibr}) is very much analogous with the
relation (\ref{(f,h)_g}). Of course, we may consider both deformations
of the ordinary differential calculus on the superspace simultaneously.
In a sense, $\eta$ is the odd counterpart of $t$ in (\ref{x-dx}).

\section{Generalized gauge theory on superspace}
\setcounter{equation}{0}
We consider again the superspace differential calculus introduced in
the preceeding section. Let $\psi$ transform under the action of a
(super) group $G$ according to $\psi \mapsto \psi'= U \psi$. With
respect to local transformations on the superspace, an exterior
covariant derivative can be defined in the usual way as
\begin{eqnarray}
                 D\psi := d\psi + A \, \psi
\end{eqnarray}
with a connection 1-form $A$. It is indeed covariant, i.e.
$D'\psi'=\hat{U} \, D\psi$, if
\begin{eqnarray}
             A' = \hat{U} \, A \, U^{-1} - dU \, U^{-1}  \; .
\end{eqnarray}
Inserting the decomposition
\begin{eqnarray}
    A = d\eta \, \alpha + dx^i \, A_i + d\xi_i \, \Lambda^i
\end{eqnarray}
we find
\begin{eqnarray}
   A'_i = U \, A_i \, U^{-1} - (\partial_i U) \, U^{-1}
            \quad , \quad
   \Lambda'^i = \hat{U} \, \Lambda^i \, U^{-1} - (\zeta^i U)
                  \, U^{-1}
\end{eqnarray}
and
\begin{eqnarray}
  \mu' = \hat{U} \, \mu \, U^{-1} - (\partial_\eta U) \, U^{-1}
             \quad , \quad
  \mu := \alpha + \hat{A}_i \, \Lambda^i - \zeta^i A_i  \; .
\end{eqnarray}
In order to read off gauge covariant components from covariant
(generalized) differential forms, we need the following covariantized
differentials (cf also section 4),
\begin{eqnarray}
   Dx^i:=dx^i - d\eta \, \Lambda^i     \quad , \quad
   D\xi_i := d\xi_i - d\eta \, \hat{A}_i    \; .
\end{eqnarray}
Their transformation rule is
\begin{eqnarray}
 D'x^i = \hat{U} \, Dx^i \, U^{-1}     \quad , \quad
 D'\xi_i = \hat{U} \, D\xi_i \, \hat{U}^{-1}  \; .
\end{eqnarray}
Now we find
\begin{eqnarray}            \label{Dpsi}
 D\psi = d\eta \, (D_\eta \psi + \Gamma^i D_i \psi)
         + Dx^i \, D_i \psi + D\xi_i \, \Gamma^i \psi
\end{eqnarray}
where
\begin{eqnarray}
 D_\eta :=  \partial_\eta + \mu  \quad , \quad
 D_i :=  \partial_i + A_i        \quad , \quad
 \Gamma^i := \zeta^i + \Lambda^i    \; .
\end{eqnarray}
The operator $\Gamma^i D_i$ (the covariantized $\Delta$) which
appears in (\ref{Dpsi}) is a {\em generalization of the Dirac
operator}. If a metric tensor $g^{ij}$ is given and $\zeta^i U = 0$,
we can choose $\Lambda^i = g^{ij} \xi_j = \xi^i$ so that $\Gamma^i =
\zeta^i + \xi^i$ and
\begin{eqnarray}
   \Gamma^i \Gamma^j + \Gamma^j \Gamma^i = 2 \, g^{ij}
\end{eqnarray}
which is the Clifford algebra relation. In this case, $\Gamma^i D_i$
is indeed the Dirac operator.

More generally, we have the following relations between transformation
properties and exterior covariant derivatives,
\begin{eqnarray}
\begin{array}{c@{\; \mapsto \;}c@{\quad \Rightarrow \quad}c@{\,=\,}l
 @{\; \, \mapsto \;}c}
 \psi & U\psi             & D\psi & d\psi+A \, \psi
                          & \hat{U} \, D \psi                     \\
 \psi & \hat{U}\psi       & D\psi & d\psi-\hat{A} \, \psi
                          & U \, D \psi                           \\
 \psi & \psi U^{-1}       & D\psi & d\psi-\hat{\psi} \, A
                          & D \psi \, U^{-1}                      \\
 \psi & \psi \hat{U}^{-1} & D\psi & d\psi+\hat{\psi} \, \hat{A}
                          & \; \, D\psi \, \hat{U}^{-1}      \; .
\end{array}
\end{eqnarray}
The curvature 2-form of the connection $A$ is given by
\begin{eqnarray}
                F := dA - \hat{A} \, A  \; .
\end{eqnarray}
We will leave the further investigation of this calculus to
a separate work.

\vskip.5cm
\noindent
{\bf Acknowledgments} \\
A. D. is grateful to the Heraeus-Foundation for financial support.
F. M.-H. would like to thank the Deutsche Forschungsgemeinschaft
for a travel allowance.


\end{document}